\documentclass[11pt]{article}
\usepackage{Blois,epsfig}
\usepackage{wrapfig}

\def\be{\begin{equation}}
\def\ee{\end{equation}}
\def\bea{\begin{eqnarray}}
\def\eea{\end{eqnarray}}

\begin{document}
\vspace*{2cm}
\begin{center}
\Large{\textbf{XIth International Conference on\\ Elastic and Diffractive Scattering\\ Ch\^{a}teau de Blois, France, May 15 - 20, 2005}}
\end{center}

\vspace*{2cm}
\title{DEEPLY VIRTUAL COMPTON SCATTERING AND PROMPT PHOTON PRODUCTION AT
HERA}

\author{L. FAVART}

\address{I.I.H.E., Universit\'e Libre de Bruxelles, Belgium}

\maketitle\abstracts{
Recent results on the Deeply Virtual Compton Scattering (DVCS) and
prompt photon productions from H1 and ZEUS experiments on the $ep$ 
collider HERA are presented. 
A new DVCS cross section measurements of the H1 Collaboration,
for photon virtualities $Q^2>4$ GeV$^2$ and photon-proton c.m.s. 
energy $30<W<140$ GeV, are
discussed and compared to NLO QCD calculations encoding Generalized
Parton Distributions (GPDs) and to Colour Dipole model predictions. 
For the first time the cross section dependence is reported
on the momentum transfer squared at the proton vertex, $t$.
Prompt photon production in deep inelastic scattering and 
photoproduction are presented both in the inclusive case and in the 
presence of a jet. The results are compared to NLO QCD predictions.
}

\section{Introduction}
This paper presents new and recent results on two processes 
allowing for test of QCD in the perturbative regime (pQCD) and 
containing a measured photon in the final state.\\
1) Deeply Virtual Compton Scattering (DVCS), $ep \to e\gamma p$,
sketched in Fig.~\ref{fig:diag}a,
consists of the hard diffractive
scattering of a virtual photon off a proton.
The interest of the DVCS process resides in the
particular insight it gives to the applicability of
perturbative Quantum Chromo Dynamics (QCD) in the field of
diffractive interactions and to the nucleon partonic structure.\\
2) Prompt photons in the final state of high energy collisions 
(Fig.~\ref{fig:diag}b and c) allow for a
detailed study of pQCD and of the hadronic structure of the incoming particles.
The term ``prompt'' refers to photons which are radiated directly from
the partons of the hard interaction.
In contrast to jets, photons are not affected by hadronisation,
resulting in a more direct correspondence to the underlying partonic 
event structure.

\begin{figure}[htb]
 \begin{center}
  \vspace*{-0.3cm}
  \epsfig{figure=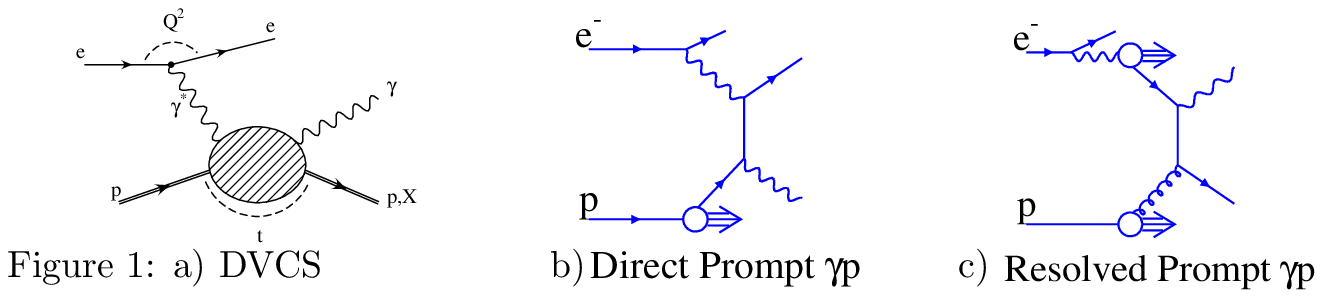,width=0.88\textwidth}
  \vspace*{-1.0cm}
  \addtocounter{figure}{1}
  \label{fig:diag}
 \end{center}
\end{figure}

\section{Deeply Virtual Compton Scattering}

DVCS cross section measurements~\cite{h1dvcs97,zeusdvcs} at HERA, 
similar to diffractive vector meson electroproduction~\cite{alessia} 
but with a real 
photon replacing the final state vector meson,  
become an important source of information to study the partons, in
particular gluons,
inside the proton for nonforward kinematics and its relation with
the forward one.
In hard exclusive production 
the proton structure has to be encoded in a generalized form
(Generalised Parton Distributions or GPDs)
to include the difference of longitudinal momentum fractions of the two 
partons, $\xi$ and transverse momentum exchange at the proton vertex. 
\\

 This paper presents a measurement of 
DVCS cross section based on 46.5$pb^{-1}$ of data collected with the H1
detector at HERA in years 1996 to 2000~\cite{h1dvcs00}.
The cross section is presented as a function of $Q^2$, $W$ and $t$.


\subsection{Data Analysis}

At the present small values of $|t|$ the reaction
$e p \rightarrow e \gamma p\,$
is dominated by the purely electromagnetic
Bethe-Heitler (BH) process whose cross section,
depending only on QED calculations and proton elastic form factors, is
precisely known and therefore can be subtracted.
To enhance the ratio of selected DVCS
events to BH events the outgoing photon is selected in the
forward, or outgoing proton, region with transverse momentum larger than
$1\,{\rm GeV}$. Large values of
the incoming photon virtuality $Q^2$ are selected by detecting the
scattered
electron in the backward calorimeter with energy larger than 
$15\,{\rm GeV}$. The outgoing proton escapes down the
beam-pipe in the forward direction.
In order to reject inelastic and proton dissociation events,
no further cluster in the calorimeters with energy above
noise level is allowed and an absence of activity
in forward detectors is required.
\\


The selected DVCS sample contains 1243 events. To extract the cross
section, the data are corrected for
detector acceptance and initial state radiation using the Monte Carlo
simulation program MILOU~\cite{Milou}.
The measured $e p \rightarrow e \gamma p$ cross
section is converted to the $\gamma^* p \rightarrow \gamma p$
cross section using equivalent photon approximation.

\subsection{Results}

\begin{wrapfigure}[15]{L}{0.55\linewidth}
 \begin{center}
  \epsfig{figure=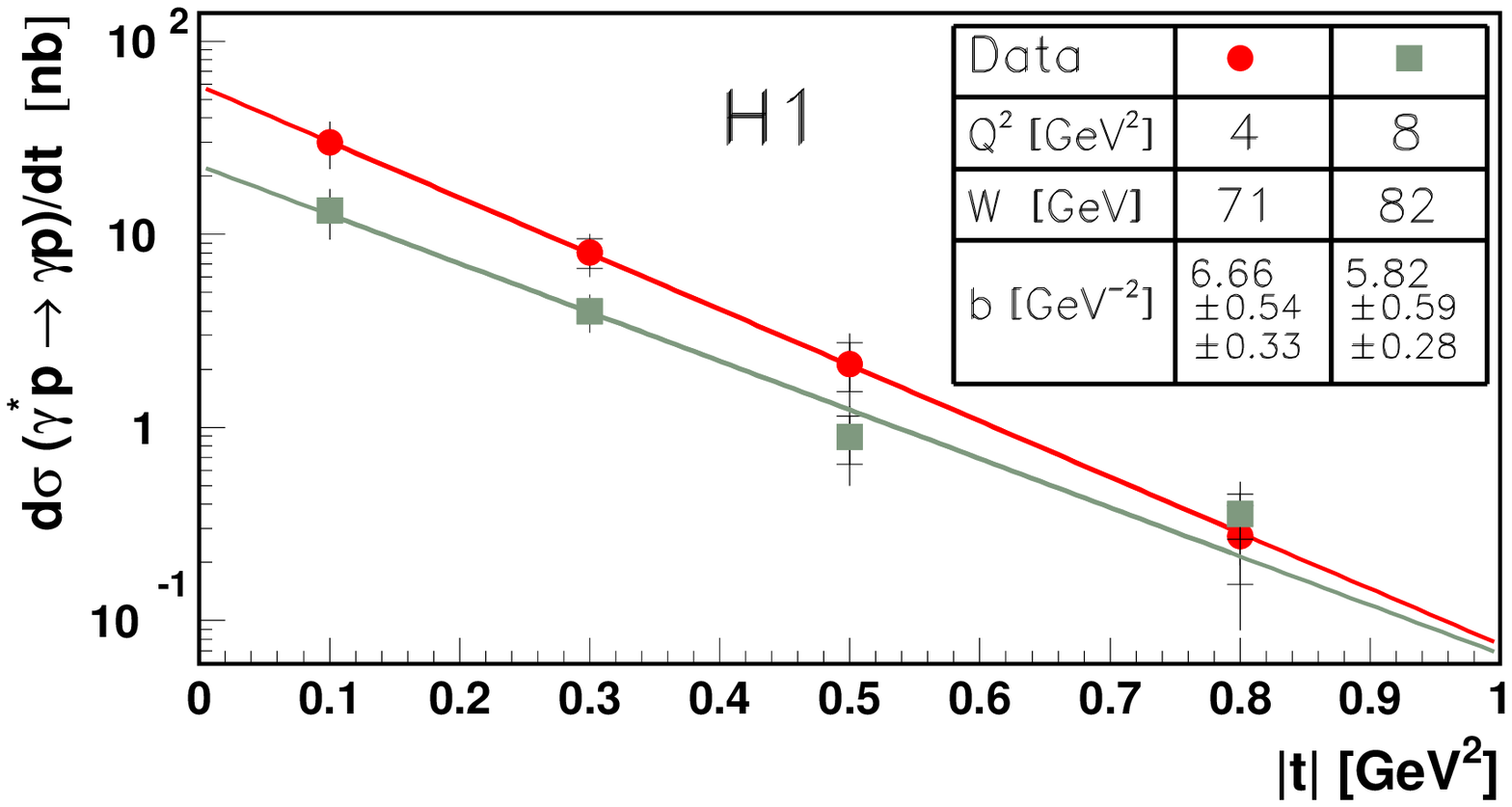,width=0.55\textwidth}
 \end{center}
 \vspace*{-0.4cm}
 \caption{\sl The cross section $\gamma^* p \rightarrow \gamma p$
  differential in $t$, for $Q^2=4$~GeV$^2$ and $Q^2=8$~GeV$^2$.
  The inner error bars represent the statistical and the
  full error bars the quadratic sum of the statistical and systematic
  uncertainties.}
 \label{fig:dsdt}
 \vspace*{-0.4cm}
\end{wrapfigure}

For triggering reasons, the cross sections are measured separately
in 1996-1997 and
1999-2000, covering different $Q^2$ ranges, and are then combined.
The differential cross section in $t$ is measured at two different $Q^2$
values as shown in Fig.~\ref{fig:dsdt}. The  $t$ dependence is
parametrised as $e^{-b|t|}$. Combining the two data sets, the
$t$ slope is measured to be 
$b = 6.02\pm 0.35{\rm(stat)} \pm 0.39{\rm(sys)}$ GeV$^{-2}$ 
for $Q^2 = 8$ GeV$^2$ and $W = 82$ GeV.\\

\begin{figure}[htb]
 \begin{center}
  \epsfig{figure=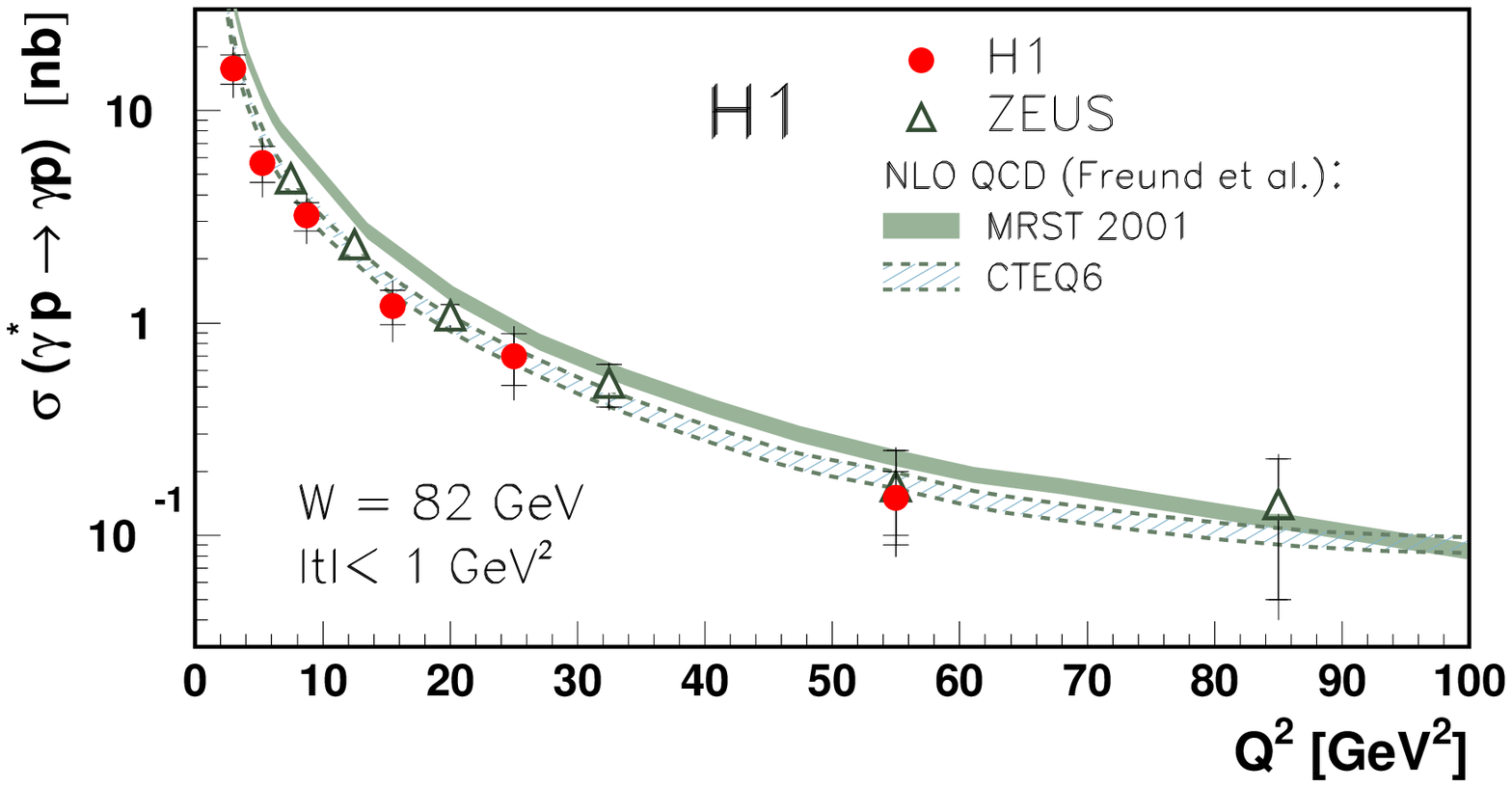,width=0.49\textwidth}
  \epsfig{figure=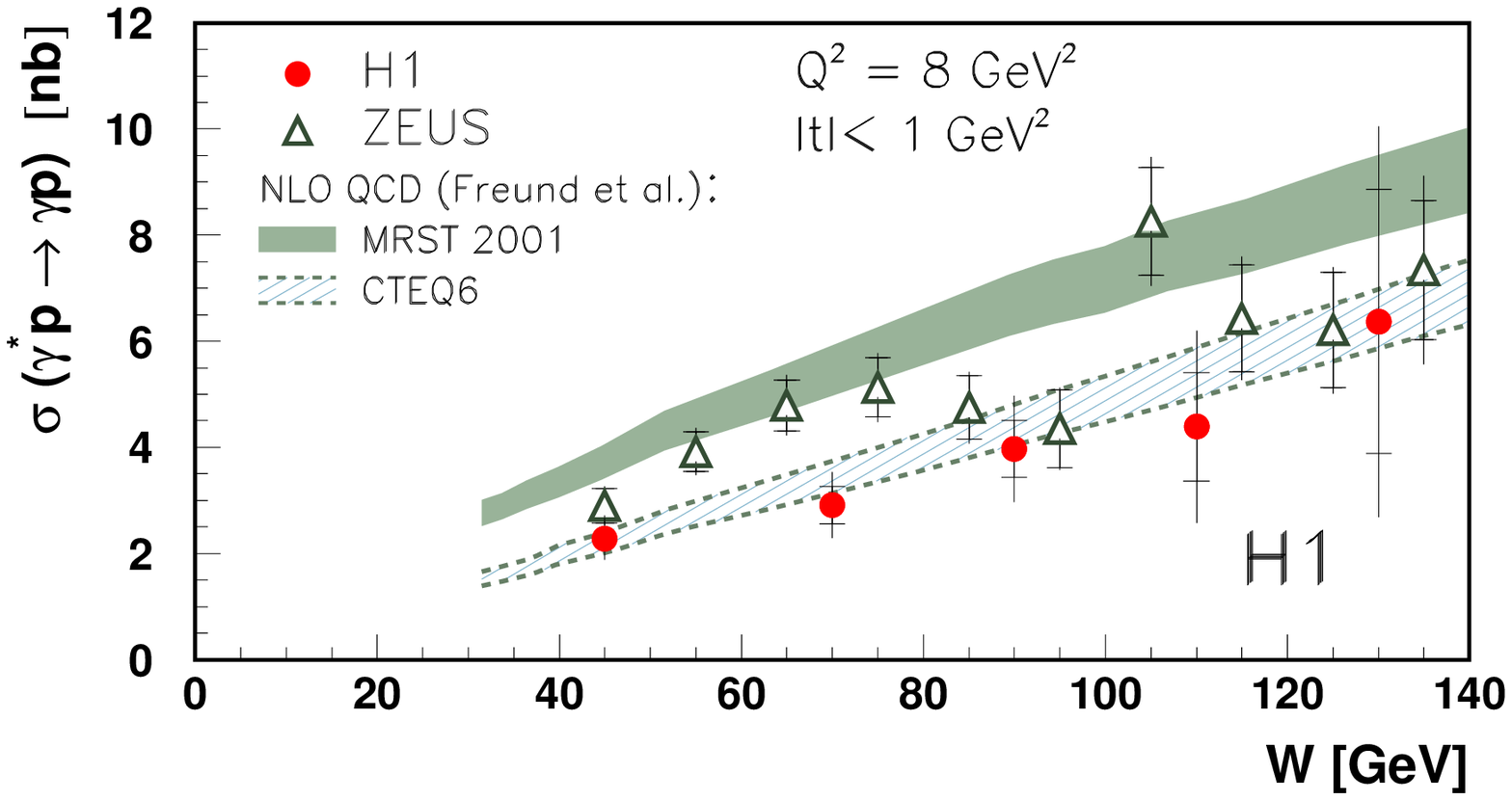,width=0.49\textwidth}
  \vspace*{-0.4cm}
 \end{center}
 \caption{\sl The $\gamma^* p \rightarrow  \gamma p$ cross section
  as a function of $Q^2$ for $W=82$ GeV (left) and 
  as a function of $W$ for $ Q^2=8$ GeV$^2$ (right).
  The H1 measurement is shown together with the results of
  ZEUS and NLO QCD predictions based on MRST 2001 
  and CTEQ6 PDFs.
  The band associated with each prediction corresponds to the
  uncertainty on the measured $t$-slope.
  }
 \label{fig:sqcd}
 \vspace*{-0.4cm}
\end{figure}

Fig.~\ref{fig:sqcd} shows the cross section as a function
of $Q^2$ for $W=82$~GeV and as a function of $W$ for $Q^2=8$~GeV$^2$.
Fitting the $Q^2$ dependence with a form
$(1/Q2)^n$ gives $n = 1.54 \pm 0.09 {\rm(stat)} \pm 0.04{\rm(sys)}$. 
Fitting the W dependence with a form $W^\delta$ gives 
$\delta=0.77\pm0.23\pm0.19$.
No $Q^2$ dependence is observed for $\delta$. The steep rise of the cross
section with $W$ indicates the presence of a hard
scattering process. The value of $\delta$ is similar to that measured in
exclusive $J/\Psi$ production. 
The new H1 measurement is
found to be in agreement with the published ZEUS results~\cite{zeusdvcs}.
Fig. \ref{fig:sqcd} also compares the measurements with QCD predictions 
calculated at NLO by Freund and McDermott~\cite{Freund:2002qf}.
In this prediction, the classic PDF $q(x,\mu^2)$ of MRST2001
and CTEQ6 are used in the DGLAP region
($|x|>\xi$) such that $\cal H$, which is the only important GPD at small
$x$ is given at the scale $\mu$ by:
${\cal H}^q(x,\xi,t;\mu^2)=q(x;\mu^2) \, e^{-b|t|}$ for quark singlet
and
$ {\cal H}^g(x,\xi,t;\mu^2)=x\ g(x;\mu^2) \, e^{-b|t|}$ for gluons,
i.e. independent of the skewing parameter $\xi$,
the skewing as the $Q^2$ dependences being generated dynamically (i.e.\ no
intrinsic skewing).
In the ERBL region ($|x| < \xi$), these parametrisations
have to be modified, ensuring a smooth continuation to the DGLAP 
region~\cite{Freund:2002qf}.
The theoretical estimates agree well with the data for both shape and
absolute normalisation. The uncertainty in the normalization for the
theory is significantly reduced
owing to the H1 measurement of the cross section (exponential) $t$ slope; 
this uncertainty becomes smaller than
the input PDF uncertainty which is quantified comparing MRST and CTEQ PDF
set based predictions.
 Furthermore, this shows that no intrinsic skewing is needed to describe
the DVCS cross section in the small Bjorken $x$ region.  
Comparison to Colour dipole models also provide a reasonable
description of the data (see~\cite{h1dvcs00}), 
both in shape and in normalisation.
The $Q^2$ dependence is better described by the Favart-Machado
prediction~\cite{Favart:2004uv}
when DGLAP evolution of the dipole is included.

\section{Prompt Photon}

\begin{figure}
 \vskip 2.3cm
 \begin{picture}(100,100)(0,0)
 \put(-15,0){\epsfig{figure=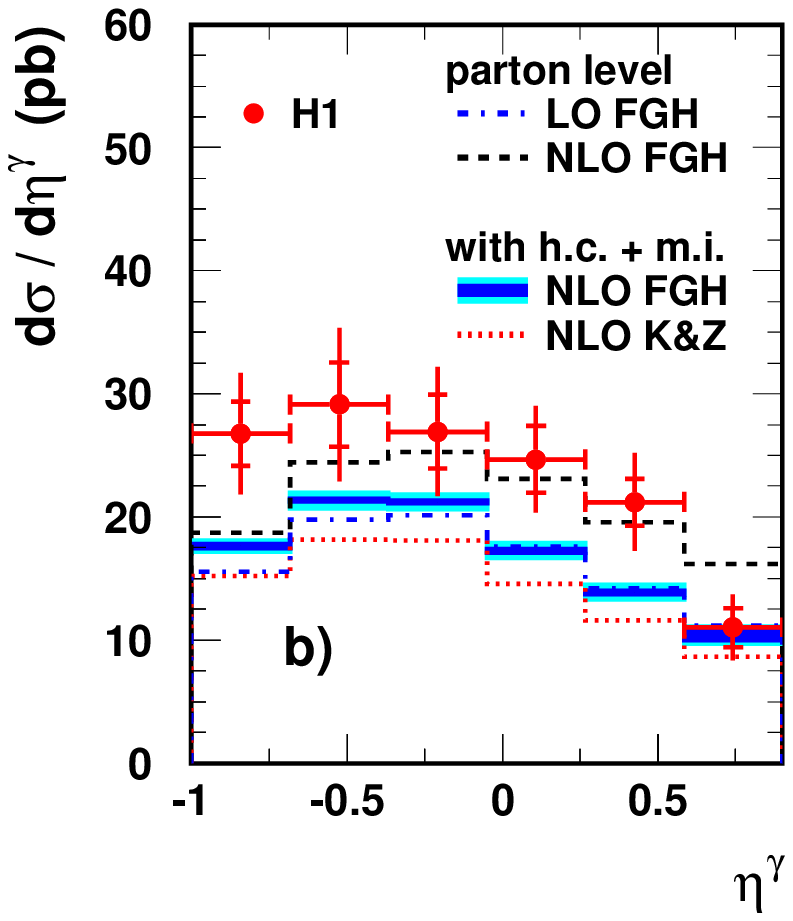,width=0.32\textwidth}}
 \put(139,0){\epsfig{figure=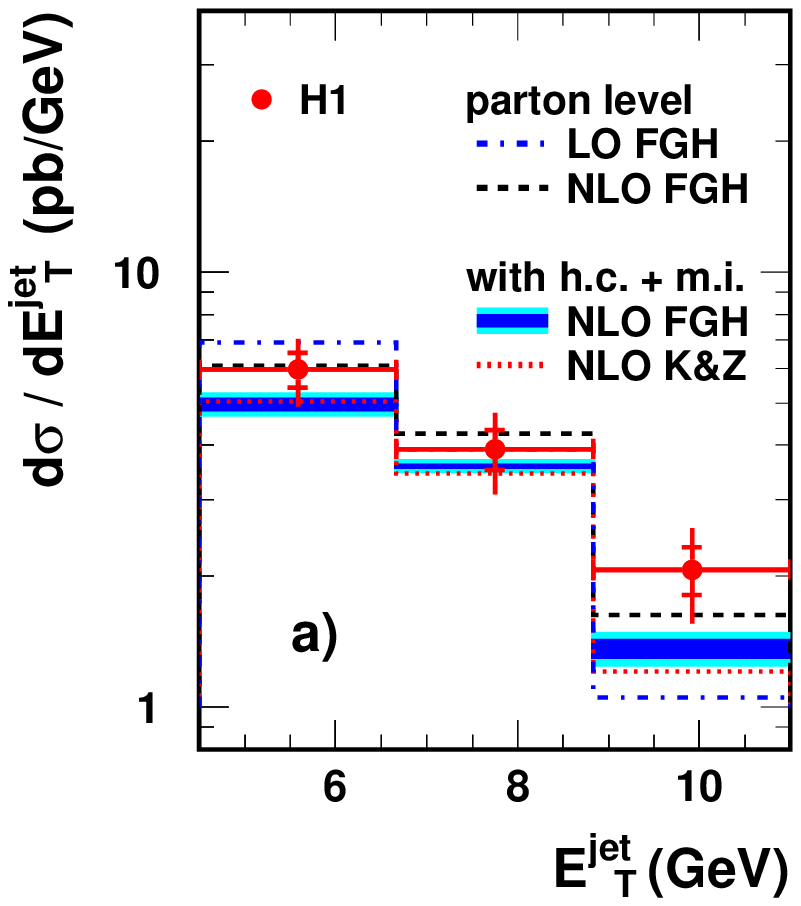,width=0.32\textwidth}}
 \put(300,13){\epsfig{figure=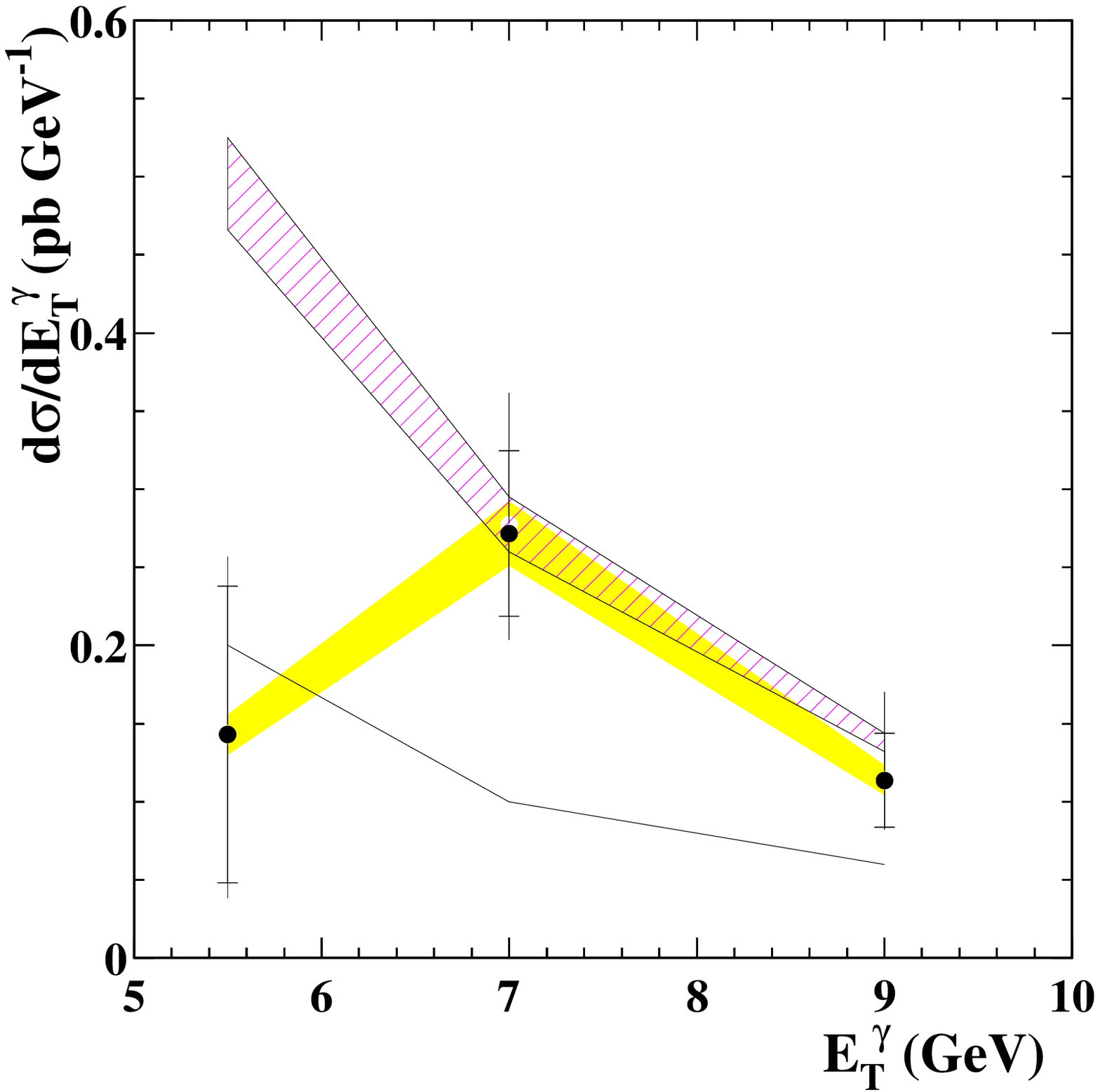,width=0.32\textwidth}}
 \end{picture}
 \vspace*{-0.6cm}
 \caption{Prompt photon differential cross sections.
  {\bf a)} in inclusive photoproduction,
  {\bf b)} in photoproduction with jet,
  {\bf c)} in DIS with jet - the hashed band represent the 
           NLO QCD prediction, the full one the energy scale
           uncertainty.
  }
 \vspace*{-0.4cm}
 \label{fig:prompt}
\end{figure}

Fig.~\ref{fig:prompt}a shows the differential inclusive prompt photon cross
section $\mathrm{d}\sigma/\mathrm{d}\eta^\gamma$ as a function
 of the pseudorapidity of the photon in photoproduction regime
($Q^2<1\,\mathrm{GeV}^2$ and $142<W<266\,\mathrm{GeV}$) by 
H1~\cite{Aktas:2004uv}. These results are compatible with ZEUS
measurement~\cite{Breitweg:1999su} (not shown). 
A comparison to NLO pQCD calculations by Fontannaz,
Guillet and Heinrich (FGH~\cite{Fontannaz:2001ek})
and Krawczyk and Zembrzuski (K\&Z~\cite{Krawczyk:2001tz}) shows a good
shape description but a normalisation $20\%-40\%$ below the data.
On Fig.~\ref{fig:prompt}b the cross section is shown differentially in 
the transverse jet energy, when an additional jet 
($E_t>4.5\,\mathrm{GeV}$ and $-1<\eta^{jet}<2.3$) is required.
Here both NLO calculations~\cite{Fontannaz:2001nq,Zembrzuski:2003nu} are
consistent with the data in most bins. 
The hadronic and multiple interactions corrections improve the description 
of the data only in some regions.
The ZEUS Collaboration~\cite{Chekanov:2004wr} has measured the inclusive
prompt photon in DIS ($Q^2>35\,\mathrm{GeV}^2$) with a jet 
($E_{T}^{jet}>6\,\mathrm{GeV}$ and $-1.5<\eta^{jet}<1.8$), see
Fig.~\ref{fig:prompt}c.
A pQCD calculation~\cite{Gehrmann-DeRidder:2000ce} to order
$\mathcal{O}(\alpha^3\alpha_s^1)$ on parton level
describes the normalisation except at low $E_T$ and in the
more forward (proton beam) direction.
The PYTHIA and HERWIG Monte Carlo predictions undershoot the data
in all cases (photoproduction and DIS),
see~\cite{Aktas:2004uv,Chekanov:2004wr} for more details.

\begin{flushright}
{\small
The author is supported by the {\it Fonds National \\
de la Recherche Scientifique} of Belgium.}
\end{flushright}
\end{document}